\documentclass[useAMS,onecolumn]{mn2e}
\usepackage{psfig}
\newif\ifAMStwofonts

\title[Millisecond and Binary Pulsars as Nature's Frequency Standards]
{Millisecond and Binary Pulsars as Nature's Frequency Standards. \\
{\LARGE III. Fourier Analysis and Spectral Sensitivity of Timing Observations 
             to Low-Frequency Noise}}
\author[Sergei M. Kopeikin and Vladimir A. Potapov]
       {Sergei M. Kopeikin $^1$
       and Vladimir A. Potapov $^2$\\ \\
      $^1$ Deptartment of Physics \& Astronomy, University of Missouri-Columbia, Columbia, MO 65211, USA \\
      $^2$ PRAO, ASC FIAN, Leninskii Prospect, 53, Moscow 117924, Russia}
\date{Accepted.............200... ;  
      Received ............200... ; 
      in original form ...........200... }
\pubyear{1999}

\def\LaTeX{L\kern-.36em\raise.3ex\hbox{a}\kern-.15em
    T\kern-.1667em\lower.7ex\hbox{E}\kern-.125emX}

\begin{document}
\Large
\label{firstpage}

\maketitle

\begin{abstract}
Millisecond and binary pulsars are the most stable natural
frequency standards which allow the introduction of modified versions of 
universal and
ephemeris time scales based on the intrinsic rotation of
pulsar and on its orbital motion around the barycentre of a binary system. 
The measured stability of these time scales depends on numerous physical
phenomena which
affect the rotational and orbital motion of the pulsar and observer on the 
Earth, perturb the 
propagation of electromagnetic pulses from the pulsar to the observer, and bring 
about random
fluctuations in the rate of time measured by an atomic clock used as a primary time reference
in timing observations. Over long time
intervals the main reason for the instability of the pulsar time scales is 
the presence of correlated, 
low-frequency timing noise in residuals of times of arrivals (TOA) of electromagnetic signals
from the pulsar
which has both astrophysical and geophysical
origins. Hence, the timing noise can carry important physical
information about the interstellar medium, the interior structure of the pulsar,
stochastic gravitational waves coming from the early universe and/or binary stars, etc. 
Each specific type of  low-frequency noise can be 
described
in a framework of the power law spectrum model. Although data processing of
pulsar timing observations in the time domain seems to be the most imformative, 
it is also
important to know to which
spectral bands single and binary pulsars, considered as detectors 
of the low-frequency noise signal, are the most sensitive. A solution to this
problem may be reached only if the parallel processing of timing data in the frequency
domain is fulfilled. This requires a development of the Fourier analysis
technique for an adequate interpretation of data contaminated by the correlated
noise with the noise spectrum which is divergent at low frequencies. 
The given problem is examined in the present article.   
\end{abstract}

\section{Introduction}

It is well known that millisecond and binary pulsars provide an exellent means for testing
the theory of general relativity (Taylor \& Weisberg 1982, 1989, van Straten {\it et al.} 2001,  Weisberg \& Taylor 2003,  Kramer {\it et al.} 2003). Pulsar timing is a powerful tool for exploring the 
structure of the interstellar medium (Rickett 1990,
1996, Shishov 2002) and for investigating the interior
of neutron stars (Cordes \& Greenstein 1981, Kaspi {\it et al.} 1994)
as well as setting an upper limit on the energy density of 
gravitational radiation produced in the early universe and/or by the orbital motion of binaries (Kaspi , Thorsett \& Dewey 1996,
McHugh {\it et al.} 1997, Kopeikin 1997a, Kopeikin 1999a, Kopeikin \& Potapov 2000, Lommen 2002, Jaffe and Backer 2003). 
Physical process of pulsar's intrinsic rotation around its axis has been 
proposed and used (Backer {\it et al.} 1982,  Il'in {\it et al.} 1986, 
Rawley {\it et al.} 1987, Guinot \& Petit 1991, Matsakis {\it et al.} 1997) as a 
new time reference (PT scale), analogous to the
universal time (UT) in time-keeping metrology. The UT is based on observation of the Earth's diurnal rotation. The rate of the atomic time scale UTC is regulated by introduction of leap seconds to reproduce the rate of UT and eliminate the secular difference between the uniform atomic time scale TAI and UT (Kovalevsky and Seidelmann 2004). Quite recently, a
new astronomical time reference (BPT scale) spread out over extremely long 
intervals (approaching 50-100 years) has been suggested 
(Petit \& Tavella 1996) and its stability have been carefully examined by Ilyasov {\it et al.} (1998) and by Kopeikin (1999).
BPT scale is derived from the orbital motion of a pulsar in a binary system and
represents an analogue of ephemeris time (ET) in classical astronomy as
based on the orbital motion of
the Earth around the Sun (or the Moon around the Earth) and 
introduced
by Newcomb (1898) at the end of the last century. 

An adequate analysis of the timing data requires a deeper understanding of the nature of
the noise process present in pulsar timing residuals. As soon as the
autocovariance function of the noise process is known, analysis in the time
domain becomes possible. Time domain analysis is the most informative since
the observed stochastic process is not stationary and includes 
a non-stationary component which contribution to the parameters' estimates can be important (Groth 1975, Cordes 1978, 1980; Kopeikin
1997b). Because pulsar timing observations are conducted over relatively
long time intervals, the white noise from errors in measuring TOA of pulsar's 
pulses is suppressed by the presence of a number of 
correlated, low-frequency (``red") noise processes having different spectra and
intensities. Henceforth, we are mainly interested in analysing the
red noise. 

A simple, but realistic mathematical model for the red noise has
been developed by Kopeikin (1997b). The model is based on the shot
noise approximation with autocovariance
function depending on both stationary and non-stationary components. The given model elucidates a rather
remarkable fact (Kopeikin 1999b), namely, that the timing 
residuals as well as variances of some spin-down and all orbital parameters are not
affected by the non-stationary component of the red noise at all because its contribution is filtered out by the fitting procedure based on the least square method. 
This observation justifies and puts on a firm ground the  
Fourier analysis of TOA residuals and variances of fitting parameters 
in the frequency domain which is normally done under implicit assumption that the non-stationary component of the autocovariance function is either absent or unimportant. We conclude that the Fourier analysis gives unbiased information about the noise process itself and permits us to reveal which
frequency harmonics in the spectral expansion of the stochastic process of the 
pulsar timing observations are the most substantial.      

Any low-frequency noise can be approximately 
characterized by the power-law spectrum $S(f) \sim
f^{-n}$ where the spectral (integer) index $n \geq 1$. 
It is obvious that the spectrum has a singularity
at zero frequency. Hence, the energy of TOA residuals comprised in
such noise should formally have infinite value because the integral over all frequencies
from zero to infinity taken from $S(f) \sim
f^{-n}$ is divergent. Clearly, this has no
physical meaning, and one has to resort to special mathematical methods
in order to avoid this illicit divergency. There are two standard mathematical methods for 
curing this flaw. The first method consists in the analytical continuation of the spectrum by formally changing it form from
$S(f)\sim f^{-n}$ to $S(f,A)\sim f^{-(n+A)}$ where $A$ is a purely imaginary 
parameter different from zero. Such a model of the analitically continued spectrum gives convergent 
integrals which coincide everywhere
on the real axis  with the integrals from the original spectrum $S(f)\sim f^{-n}$, exept 
for the point $f=0$. In order to
prescribe a physical meaning to such integrals, we have to expand them in
a Laurent series with respect to the parameter $A$ and take the finite part
of the expansion for $A=0$. Such procedure has been used, in particular, 
by Kopeikin (1997a) for regularization of the autocovariance function of stochastic noise of the primordial 
gravitational wave background having formally divergent spectrum $S(f) \sim f^{-5}$ (Sazhin 1978, Detweiler 1979, Mashhoon 1982, 1985,  Bertotti {\it et al.} 1983, Mashhoon \& Seitz 1991)  and for setting an upper limit on the energy density of this background noise in the ultra-low frequency domain.

The procedure of analytic
continuation of divergent integrals is mathematically rigorous and represents
theoretically a powerful tool which gives well-defined and self-consistent mathematical
results (Gel'fand \& Shilov 1964). However, for researchers doing numerical computations the second method of regularization of singular spectra is preferable and more useful practically. It is based on the 
infrared cutoff of all divergent integrals at the frequency
$f_{\rm cutoff}=\varepsilon$, where $\varepsilon\ll 1$ is constant. The infrared cutoff technique requires corresponding modification of the power-law spectrum of
the red noise in order to eliminate the dependence of the
fitting parameters and the autocovariance function of the noise on the {\it ad hoc} introduced frequency cutoff $\varepsilon$. 
We shall prove that this modification consists of subtracting from the original power-law spectrum the infinite series
of the frequency-dependent Dirac's delta function and its derivatives with numerical coefficients which are functions of
the cutoff frequency $\varepsilon$. Such modification of the spectrum
preserves the structure of the most, generally accepted, autocovariance functions and,
as a consequence, does not change results of the numerical processing of signals
in time domain. In addition, the infinite delta-function dependent part of the modified spectrum is rapidly convergent and accounting for contribution of few first terms is sufficient in practice. This second method of 
regulariazation of red spectra being divergent in the infrared frequency domain is worked out in the present paper.  

In what follows, it is more convenient for mathematical purposes to express results in terms of units in which 
frequency and time are dimensionless. For instance, in binary pulsars, it is 
preferable for symbolic calculations 
to measure time in units of orbital frequency, $n_b=2\pi/P_b$, where
$P_b$ is the orbital period of the binary system. In these units, frequency $f$ is measured in
terms of $1/P_b$, and dimensionless time is the pulsar's mean orbital anomaly
$u=n_b t$ where $t$ is time measured in SI units. 

\section{Regularized Spectrum of Low-Frequency Noise}

Any gaussian low-frequency noise is completely characterized by the 
autocovariance
function which describes the correlation between two values of stochastic process
separated by the arbitrary time interval $u=n_b(t_2-t_1)$. The autocovariance
function usually consists of two algebraically independent parts characterizing
separately a stationary, $R^-(u)$, and a non-stationary, $R^+(u)$, 
component of the noise. Complete expressions of the
autocovariance functions for
various examples of low-frequency noise have been derived by 
Kopeikin (1997b), where the shot noise approximation of a stochastic process has
been used. Although both stationary and non-stationary components of the
autocovariance function are important for a comprehensive treatment of observations, we are dealing in the present paper only with the stationary part since the non-stationary component of the red noise is filtered out by the least square method (Kopeikin 1999b).

The stationary part of the authocovariance function can be transformed into the spectral density of noise,
$S(f)$, by means of the Wiener-Khintchine theorem
\vspace{0.3 cm}
\begin{equation}
\label{aa}
R^-(u)=2\int_0^{\infty}S(f)\cos(2\pi f u)df,
\end{equation}
where $u$ is the dimensionless time and $f$ is the dimensionless Fourier frequency.
If the (constant) intensity of noise is denoted by $h_n$ then the
autocovariance function of
low-frequency (correlated) noise is determined by the expression (Kopeikin 1997b)
\vspace{0.3 cm}
\begin{equation}
\label{gr}
\left\{\begin{array}{ll}C_n h_n |u|^{n-1}, &\mbox{$n=2,4,6,...$,
random walk noise}\\ \\
C_n h_n u^{n-1} \ln|u|, 
& \mbox{$n=1,3,5,...$, flicker noise}
\end{array}\right. 
\end{equation} 
where $C_n$ is a numerical constant of normalization.   
 
Functions, which
might be appropriate candidates for the spectrum of noise procesess with the
foregoing autocovariance functions, are $S(f)\sim h_n f^{-n}$ where $n$ is
integer. However,
integrals (\ref{aa}) from such power-law functions are divergent because of
non-physical singularity of the spectrum at zero frequency. For this reason, a regularization
technique should be used because we don't usually know the asymptotic low-frequency
behaviour of the spectrum and can not treat the extremaly low-frequency domain adequately. 

To develop the regularization method the power spectrum $S(f)$ of the noise should be defined 
using the truncated Fourier transform of the autocovariance function with the lower cutoff frequency
$f_{\rm cutoff}=\varepsilon$.  
Specifically, we require that the 
truncated cosine Fourier transform of $S(f)$ must give the stationary part 
of the original autocovariance functions (\ref{gr}) without any 
additional contributions which might come up from the cutoff-frequency lower limit of the integral and bias the estimates of the fitting parameters. Let us guess that this requirement can be satisfied with the specific model of the red-noise spectrum $S(f)$ 
represented by the formula
\vspace{0.3 cm}
\begin{equation}
S(f)=\left\{\begin{array}{ll}h_n\left[\displaystyle{
\frac{1}{(2\pi f)^n}}+\displaystyle{\sum_{k=0}^\infty}B_{2k}(\varepsilon) 
\varepsilon^{2k}\delta^{(2k)}(f-\varepsilon)\right], &\mbox{if
$f\geq\varepsilon$}\\ \\
0 , & \mbox{otherwise}
\end{array}\right.
\label{1}   
\end{equation}
\vspace{0.3 cm}
where the spectral index of noise $n=1,2,...,6$, the constant parameter $h_n$ 
is the strength of noise, quantities 
$B_k(\varepsilon)$ are constant numerical coefficients being defined
later, and $\delta^{(k)}(f-\varepsilon)$ denotes the $n-th$ derivative 
with respect to $f$
of the Dirac delta-function $\delta(f-\varepsilon)$. The Dirac delta function
is defined according to the condition (Korn \& Korn 1968)
\begin{equation}
\displaystyle{\int_a^b f(x)\delta(x-X)dx}=\left\{\begin{array}{ll}0,
&\mbox{if $X<a$, or $X>b$,}\\ \\
\frac{1}{2}f(X+0),&\mbox{if $X=a$,}\\ \\
\frac{1}{2}f(X-0),&\mbox{if $X=b$,}\\ \\
\frac{1}{2}\left[f(X-0)+f(X+0)\right],&\mbox{if $a<X<b$,}
\end{array}\right.
\label{1aa}
\end{equation}
\vspace{0.3 cm}
where $f(x)$ is an arbitrary function being such that the unilateral limits $f(X-0)$
and $f(X+0)$ exist. The proposed form of the low-frequency spectrum $S(f)$ will be acceptable if the 
coefficients $B_k(\varepsilon)$ can be uniquely determined by the condition that 
the cosine Fourier transform
of $S(f)$ gives the stationary part of the autocovariance function of a corresponding
low-frequency noise shown in Eq. (\ref{gr}).  This statement is valid and its consistency can be proven by making use of straightforward mathematical calculations. 
As an example, we show how to determine the
several first
coefficients $B_k(\varepsilon)$ in the event of flicker noise in pulsar's
rotational phase 
which has the spectral
index $n=1$. Coefficients $B_k(\varepsilon)$ for noises having other 
spectral indices will be displayed in this section without proof (which is
rather easy). 

The stationary part of the autocovariance function $R^{-}(\tau)$ of the flicker noise in the
pulsar's phase is 
equal to $h_1 \pi^{-1} \log|u|$. 
According to the definition of the spectrum, accepted in the present paper, we should have
\vspace{0.3 cm}
\begin{equation}
R^{-}(u)=2\displaystyle{\int_{\varepsilon}^{\infty}}S(f) \cos(2\pi f u)df,
\label{2}
\end{equation}
\vspace{0.3 cm}
where the integral has been truncated at the cutoff frequency $\varepsilon\ll 1$ to eliminate its divergency at the point $f=0$. 
Substituting $S(f)$ from Eq. (\ref{1}) with $n=1$ 
on right hand side of Eq. (\ref{2}) and taking the integral we get
\vspace{0.3 cm}
\begin{equation}
R^{-}(u)=-\frac{h_1}{\pi}{\rm Ci}(2\pi \varepsilon |u|)+
h_1\cos(2\pi\varepsilon u)
\displaystyle{\sum_{k=0}^\infty}(-1)^k B_{2k}(\varepsilon) 
(2\pi\varepsilon u)^{2k},
\label{3}   
\end{equation}
\vspace{0.3 cm}
where ${\rm Ci}(x)$ is the cosine integral, and 
we have used the formula (Korn \& Korn 1968)
\vspace{0.3 cm}
\begin{equation}
2\displaystyle{\int_{\varepsilon}^{\infty}}\delta^{(2k)}(f-\varepsilon)
\cos(2\pi f u)df=\frac{d^{2k}}{d\varepsilon^{2k}}
\cos(2\pi\varepsilon u)=
(-1)^k (2\pi u)^{2k}\cos(2\pi\varepsilon u).
\label{4}
\end{equation}
\vspace{0.3 cm}
A Taylor expansion of the cosine integral and $\cos(2\pi\varepsilon u)$ in the 
right hand side of the expression (\ref{3}) with respect 
to small parameter $\varepsilon$ yields
\vspace{0.3 cm}
\begin{equation}
R^{-}(u)=\frac{h_1}{\pi}\left[-\log |u|-\gamma-\log(2\pi\varepsilon)\right]+
h_1\biggl\{B_0(\varepsilon)+(2\pi\varepsilon u)^2
\left[\frac{1}{4\pi}-\frac{1}{2}B_0(\varepsilon)-
B_2(\varepsilon)\right]\biggr\}+O(\varepsilon^4).
\label{5}
\end{equation}
\vspace{0.3 cm}
Since by definition $R^{-}(u)$ must be equal to $-h_1\pi^{-1}\log |u|$, we find 
from Eq. ({\ref 5})
\vspace{0.3 cm}
\begin{equation}
B_0(\varepsilon)=\frac{\gamma}{\pi}+\frac{\log(2\pi\varepsilon)}{\pi},\hspace{1.5 cm}
B_2(\varepsilon)=\frac{1}{4\pi}-\frac{\gamma}{2\pi}-\frac{\log(2\pi\varepsilon)}{2\pi},\hspace{1.5 cm} 
B_4(\varepsilon)=O(\varepsilon^4).
\label{6}
\end{equation}
\vspace{0.3 cm}
We believe that for practical purposes it is enough to account for the coefficient 
$B_0(\varepsilon)$ only,
since all other residual terms that appear are multiplied by the factor 
$2\pi\varepsilon u$,
which is expected to be negligibly small under usual circumstances, because of the smallness of the 
product $\varepsilon u$. From the formally mathematical point of view the smaller the cutoff frequency $\varepsilon$ the more exact is our approximation. However, one should take into account that the real spectrum of the low-frequency noise can have the low-frequency behaviour different from the power law and our approximation will not get better for a very small value of $\varepsilon$.  In other words 
the residual terms in the model of the spectrum under
discussion are model dependent. Had we chosen another model for the spectrum
having different behaviour as the frequency approaches zero 
the residual terms would look different. What model of the low frequency noise works better in the procedure of fitting parameters of the pulsar is an interesting question for further discussion. We emphasize however that the definition of the low-frequency noise spectrum in terms of the power-law model amended by the series of delta-functions and its derivatives makes the estimates of the fitting parameters not sensitive to the cutoff frequency and, thus, extrapolates the power-law model of the noise spectrum to lower frequencies than the power-law model of noise which does not account for the delta functions.

Proceeding in the same way for the set of other spectral indeces we obtain 
the following expressions for the power spectra of low-frequency noises:

\begin{enumerate}
\renewcommand{\theenumi}{(\arabic{enumi})}

\item \underline{Flicker noise in phase}:
\vspace{0.3 cm}
\begin{equation}
S(f)=\frac{h_1}{2\pi}\biggl\{
\frac{1}{ f}+2
\left[\gamma+\log(2\pi\varepsilon)\right]\delta(f-\varepsilon)\biggr\}+
O(\varepsilon^2).
\label{7}   
\end{equation}
\vspace{0.3 cm}

\item \underline{Random walk in phase}:
\vspace{0.3 cm}
\begin{equation}
S(f)=\frac{h_2}{4\pi^2}\biggl\{
\frac{1}{f^2}-
\frac{2}{\varepsilon}\delta(f-\varepsilon)\biggr\}+
O(\varepsilon).
\label{8}
\end{equation}
\vspace{0.3 cm}

\item \underline{Flicker noise in frequency}:
\vspace{0.3 cm}
\begin{equation}
S(f)=\frac{h_3}{8\pi^3}\biggl\{
\frac{1}{f^3}-
\frac{1}{\varepsilon^2}\delta(f-\varepsilon)
+\left[\log(2\pi\varepsilon)+\gamma-1\right]
\delta^{(2)}(f-\varepsilon)\biggr\}+
O(\varepsilon^2).
\label{9}
\end{equation}
\vspace{0.3 cm}

\item \underline{Random walk in frequency}:
\vspace{0.3 cm}
\begin{equation}
S(f)=\frac{h_4}{16\pi^4}\biggl\{
\frac{1}{f^4}-
\frac{2}{3\varepsilon^3}\delta(f-\varepsilon)
-\frac{2}{3\varepsilon}
\delta^{(2)}(f-\varepsilon)\biggr\}+
O(\varepsilon).
\label{10}
\end{equation}
\vspace{0.3 cm}

\item \underline{Flicker noise in frequency derivative}:
\vspace{0.3 cm}
\begin{equation}
S(f)=\frac{h_5}{32\pi^5}\biggl\{
\frac{1}{f^5}-
\frac{1}{2\varepsilon^4}\delta(f-\varepsilon)
-\frac{1}{4\varepsilon^2}
\delta^{(2)}(f-\varepsilon)
+\frac{1}{12}\left[\gamma+\log(2\pi\varepsilon)-\frac{1}{3}\right]
\delta^{(4)}(f-\varepsilon)\biggr\}+
O(\varepsilon^2).
\label{11}
\end{equation}
\vspace{0.3 cm}

\item \underline{Random walk in frequency derivative}:
\vspace{0.3 cm}
\begin{equation}
S(f)=\frac{h_6}{64\pi^6}\biggl\{
\frac{1}{f^6}-
\frac{2}{5\varepsilon^5}\delta(f-\varepsilon)
-\frac{2}{15\varepsilon^3}
\delta^{(2)}(f-\varepsilon)
\biggr\}+
O(\varepsilon).
\label{rww}
\end{equation}
\end{enumerate}
\vspace{0.3 cm}
Note that the coefficient $B_4(\varepsilon)\equiv 0$ in 
the expression (\ref{rww}) for the spectrum of the random walk in the
pulsar's frequency derivative.
  
The expressions given above indicate that there is a strong concentration of the
infinite energy of the noise at
the lower cutoff frequency. As we have stressed already it is a specific
feature of the chosen model of the spectrum which appears because we do not 
know the  real behaviour
of the spectrum while the Fourier frequency is approaching zero. Another remark is that 
a formal integration of any of the foregoing spectra with spectral indices $n\ge 2$ with respect
to frequency from $f=\varepsilon$ to infinity 
(which may be naively treated as the energy being stored in TOA residuals) 
gives zero value which may look surprising \footnote{A simplest way to see this result is to notice that such integration corresponds to the value of the autocovariance function $R(u)$ taken at the point $u=0$.}. However, 
it is worth noting that the whole amount of energy presents in TOA residuals can 
be calculated
only after multiplication of the spectrum by the filter function of the fitting procedure (see section 6
below for more detail).
Therefore, calculation of the total energy of residuals is more complicated and
always gives a positive numerical value as expected. Similar arguments 
can be used in
calculating variances of the fitting parameters. For example, the calculation of
variances of the first several spin-down parameters in the Fourier frequency
domain may give a negative numerical value for the variance (Kopeikin 1999b)
which is physically meaningless. The paradox is solved if we take into account the
contribution of the non-stationary part of the noise which always makes variances
of the parameters numerically positive \footnote{For more detail see (Kopeikin 1999b)}.
 
Pulsar timing observations can be used for the estimation of the strength and spectrum
of low frequency noise present in TOA residuals. For this reason, the development
of practically useful estimators of spectrum of noise are
required. We are not going to consider in the present paper the question 
of how to construct the best possible estimators. This subject has been 
significantly explored by a number of other
researches (see, for instance, the papers of Deeter \& Boynton
(1982), Deeter(1984), Taylor 1991, Matsakis {\it al.} 1997, Scott {\it et al.} 2003). Our purpose is to 
study the spectral dependence of TOA residuals and the variances of 
fitting parameters which are used in real practice. In order to put our approach on a systematic basis 
connected to our previous works let us describe, first of all, the timing model we are 
dealing with.  

\section{Timing Model}

We consider a simplified, but still realistic model of arrival time
measurements of pulses from a pulsar in a binary system. It
is assumed that the orbit is circular, and
the pulsar rotates around its own axis with angular frequency $\nu _{p}$
which slows down due to the electromagnetic (and other) energy losses. It
also takes into account that the orbital frequency of the binary system, $%
n_{b},$ and its projected semimajor axis, $x,$ have a
secular drift caused by radial acceleration of the
binary (Damour \& Taylor 1991, Bell \& Bailes 1996), its proper motion  
in the sky (Kopeikin 1996), and the emission of gravitational waves from the binary (Peters
\& Mathews 1963, Peters 1964) bringing about the gravitational radiation 
reaction force (Damour 1983a, Grishchuk \& Kopeikin 1983). The timing model, which we work with, is described in full detail in our previous paper (Kopeikin 1999b) which particular notations we use in the present paper. Let us summarize those equations which will be useful for our anlysis.

We assume that all observations of the binary pulsar are of a similar quality
and weight.
Then one defines the timing residuals $r(t)$ as a difference between the
observed number of the pulse, ${\cal N}^{obs},$ and the number ${\cal N}%
(t,\theta ),$ predicted on the ground of our best guess to the prior unknown
parameters of timing model, divided by the pulsar's rotational
frequency $\nu $, that is\vspace{0.3 cm} 
\begin{equation}
r(t,\theta )=\frac{{\cal N}^{obs}-{\cal N}(t,\theta )}{\nu },\vspace{0.3 cm}
\label{188a}
\end{equation}\vspace{0.3 cm}
where $\theta =\{\theta _{a},a=1,2,...k\}$ denotes a set of $k$ measured
parameters \footnote{The numebr $k=14$ in the particular timing model we are dealing with (see the paper (Kopeikin 1999b) for fuller detail.} which are shown in 
Table
{\ref{tab:par}}. It is worth noting that hereafter we use for the reason of 
convinuence the time argument $u=n_b t$, that is the current orbital phase.
\begin{table*}
\begin{minipage}{140mm}
\centering
\scriptsize\caption{List of the basic functions and parameters used in the fitting 
procedure. Spin parameters $\delta {\cal N}_{0}, \delta {\nu },
\delta\stackrel{.}{\nu }, \delta\stackrel{..}{\nu }, 
\delta\stackrel{...}{\nu }, \delta\stackrel{....}{\nu },$ fit rotational 
motion of the pulsar around its own axis. Keplerian parameters $\delta{x},
\delta\sigma, \delta{n}_{b}$ fit the Keplerian orbital motion of the pulsar 
about the barycentre of the binary system. Post-Keplerian parameters 
$\delta\stackrel{.}{x}, \delta\stackrel{..}{x}, \delta\stackrel{...}{x}, 
\delta\stackrel{.}{n}_{b}, \delta\stackrel{..}{n}_{b}$ fit small observable
deviations of the pulsar's orbit from the Keplerian motion caused by the
effects of General Relativity, radial acceleartion, and proper motion
of the barycentre of the binary system with respect to observer.}
\label{tab:par}
\vspace{5 mm}
\begin{tabular}{|ll@{\hspace{5 cm}}ll|}
\hline \\ \\
Parameter &&& Fitting Function \\ \hline
&  \\ 
$\beta _{1}={{\frac{\delta {\cal N}_{0}}{\nu }}}$&&& $\psi _{1}(t)=1$ \\ 
&  \\ 
$\beta _{2}={\frac{1}{n_{b}}}{{\frac{\delta \nu }{\nu }}}$ &&& $\psi
_{2}(t)=u$ \\ 
&  \\ 
$\beta _{3}={\frac{1}{2n_{b}^{2}}}{{\frac{\delta \stackrel{.}{\nu }}{\nu }}}%
 $ &&& $\psi _{3}(t)=u^{2}$ \\ 
&  \\ 
$\beta _{4}={\frac{1}{6n_{b}^{3}}}{{\frac{\delta \stackrel{..}{\nu }}{\nu }}}%
$ &&& $\psi _{4}(t)=u^{3}$ \\ 
&  \\ 
$\beta _{5}={\frac{1}{24n_{b}^{4}}}{{\frac{\delta \stackrel{...}{\nu }}{\nu }%
}}$ &&& $\psi _{5}(t)=u^{4}$ \\ 
&  \\ 
$\beta _{6}={\frac{1}{120n_{b}^{5}}}{{\frac{\delta \stackrel{....}{\nu }}{%
\nu }}}$ &&& $\psi _{6}(t)=u^{5}$ \\ 
&  \\ 
$\beta _{7}=-\delta x\sin \sigma -\delta \sigma x\cos \sigma $ &&& $\psi
_{7}(t)=\cos u$ \\ 
&  \\ 
$\beta _{8}=-\delta x\cos \sigma +\delta \sigma x\sin \sigma $ &&& $\psi
_{8}(t)=\sin u$ \\ 
&  \\ 
$\beta _{9}={\frac{1}{n_{b}}}\left( -\delta \stackrel{.}{x}\cos \sigma
+\delta n_{b}x\sin \sigma \right) $ &&& $\psi _{9}(t)=u\sin u$ \\ 
&  \\ 
$\beta _{10}={\frac{1}{n_{b}}}\left( -\delta \stackrel{.}{x}\sin \sigma
-\delta n_{b}x\cos \sigma \right) $ &&& $\psi _{10}(t)=u\cos u,$ \\ 
&  \\ 
$\beta _{11}={\frac{1}{2n_{b}^{2}}}\left( -\delta \stackrel{..}{x}\sin
\sigma -\delta \stackrel{.}{n}_{b}x\cos \sigma \right) $ &&& $\psi
_{11}(t)=u^{2}\cos u$ \\ 
&  \\ 
$\beta _{12}={\frac{1}{2n_{b}^{2}}}\left( -\delta \stackrel{..}{x}\cos
\sigma +\delta \stackrel{.}{n}_{b}x\sin \sigma \right) $ &&& $\psi
_{12}(t)=u^{2}\sin u$ \\ 
&  \\ 
$\beta _{13}={\frac{1}{6n_{b}^{3}}}\left( -\delta \stackrel{...}{x}\cos
\sigma +\delta \stackrel{..}{n}_{b}x\sin \sigma \right) $ &&& $\psi
_{13}(t)=u^{3}\sin u$ \\ 
&  \\ 
$\beta _{14}={\frac{1}{6n_{b}^{3}}}\left( -\delta \stackrel{...}{x}\sin
\sigma -\delta \stackrel{..}{n}_{b}x\cos \sigma \right) $ &&& $\psi
_{14}(t)=u^{3}\cos u$ \\ \\
\hline  
\end{tabular}
\end{minipage}
\end{table*}
If a numerical value of the parameter $\theta _{a}$ coincides with its
true physical value $\hat{\theta}_{a}$, then the set of residuals would
represent a physically meaningful noise $\epsilon (t)$, {\it i.e. }
\begin{equation}
r(t,\hat{\theta})=\epsilon (t).
\end{equation}
In practice, however, the true values of parameters
are not attainable and we actually deal with their least square estimates $
\theta _{a}^{*}.$ Therefore, the observed residuals are fitted to the expression
which is a linear function of corrections to the estimates $\theta _{a}^{*}$
of a priori unknown true values of parameters $\hat{\theta}_{a}$. From a
Taylor expansion of the timing model we work with (Kopeikin 1999b), and the
fact that $r(t,\hat{\theta})=\epsilon (t)$ one obtains\vspace{0.3 cm} 
\begin{equation}
r(t,\theta ^{*})=\epsilon (t)-{\displaystyle \sum_{a=1}^{14}}\beta _{a}\psi _{a}(t,\theta
^{*})+O(\beta _{a}^{2}),  \label{199}
\end{equation}\vspace{0.3 cm}
where the quantities $\beta _{a}\equiv \delta \theta _{a}=\theta _{a}^{*}-%
\hat{\theta}_{a}$ are the corrections to the unknown true values of
parameters, and $\psi _{a}(t,\theta ^{*})=\left(\partial {\cal N}/
\partial \theta _{a}\right) _{\theta =\theta ^{*}}$ are basic fitting
functions of the timing model.

In the following discussion it is convenient to regard the increments $\beta _{a}$
as new parameters whose values are to be determined from the fitting
procedure. The parameters $\beta _{a}$ and fitting functions are summarized in
Table {\ref{tab:par}} with asterisks omitted and time $t$ is replaced in accordance to our notations and conventions by
the function $u=n_{b}t$ which is the value of the orbital phase taken at time $t$.
We restrict the model to $14$
parameters since in practice only the first several parameters of the model
are significant in fitting  the rotational and orbital phases of the pulsar over the
available time span of observations. 

Let us introduce auxilary functions $\Xi_a(t)$ defined according to the
formula
\vspace{0.3 cm}
\begin{equation}
\label{ert}
\Xi_a(t)=\displaystyle{\sum_{b=1}^{14}}L_{ab}^{-1}\psi_b(t),
\end{equation}
where the matrix of information 
\vspace{0.3 cm} 
\begin{equation}
L_{ab}(\Delta T)={\displaystyle \sum_{i=1}^{mN}}\psi _{a}(t_{i})\psi _{b}(t_{i}),\vspace{0.3 cm}
\label{111a}
\end{equation}\vspace{0.3 cm}
the matrix $L_{ab}^{-1}$ is its inverse, and $\Delta T=NP_{b}$ is a total
span of observational time.
Functions $\Xi(t)$ are dual  (Deeter 1984) to $\psi_a(t)$ 
because of the cross-orthonormality condition 
\vspace{0.3 cm}
\begin{equation}
\label{oik}
\displaystyle{\sum_{i=1}^{mN}}\Xi_a(t_i)\psi_b(t_i)=\delta_{ab}.
\end{equation}
Now suppose that we measure $m$ equally spaced and comparably accurate
arrival times for each orbit for a total of $N$ orbital revolutions, so we have $%
mN$ residuals $r_{i}\equiv r(t_{i}),$ $i=1,...,mN.$ Standard least squares
procedure (Bard 1974) gives the best fitting solution for the estimates
of the parameters $\beta _{a}$ \vspace{0.3 cm} 
\begin{equation}
\beta _{a}({\Delta T})={\displaystyle \sum_{i=1}^{mN}}\Xi_a(t_i)
\epsilon (t_{i}),\qquad a=1,...,14.\vspace{0.3 cm}  \label{111}
\end{equation}
\vspace{0.3 cm}  

Let the angular brackets denote an ensemble average over many stochastic
realizations of the observational procedure. Hereafter, we assume that 
the ensemble average of the noise $\epsilon(t)$ is equal to
zero. 
Hence, the mean value of any parameter $\beta_{a}$ is equal to zero as
well, {\it i.e.}
\begin{equation}
<\epsilon(t)> =0, \hspace{1 cm}\mbox{and} \hspace{1 cm} <\beta _{a}>=0.
\label{mean}
\end{equation}
The covariance matrix 
$M_{ab}\equiv $ $<\beta _{a}\beta _{b}>$ of the
parameter estimates is given by the expression \vspace{0.3 cm} 
\begin{equation}
M_{ab}({\Delta T})=
{\displaystyle \sum_{i=1}^{mN}}{\displaystyle \sum_{j=1}^{mN}}
\Xi _{a}(t_{i})\Xi
_{b}(t_{j})R(t_{i},t_{j}) ,\vspace{0.3 cm}  
\label{112}
\end{equation}\vspace{0.3 cm}
where $R(t_{i},t_{j})=$ $<\epsilon (t_{i})\epsilon (t_{j})>$ is the
autocovariance function of the stochastic process $\epsilon (t)$.
The covariance matrix is
symmetric (that is, $M_{ab}=M_{ba}$), 
elements of its main diagonal give
variations (dispersions) of measured parameters 
$\sigma_{\beta _{a}}\equiv M_{aa}$ = $<\beta_{a}^{2}>$, and the off-diagonal 
terms represent the degree of statistic covariance (correlation) between 
them. The covariance matrix consists of two additive components $M_{ab}^{+}$ and
$M_{ab}^{-}$ describing contributions from the non-stationary and
stationary parts of autocovariance function $R(t_i,t_j)$  respectively. Explicit expressions
for the matrix $M_{ab}$ can be found in the paper (Kopeikin 1999b) wherein we
have done all our calculations in the time domain. Only $M_{ab}^{-}$ admits the
transformation to the frequency domain and this Fourier transform will
be discussed in subsequent sections along with particular spectral properties of $M_{ab}^{-}$.

Subtraction of the adopted model from the observational data leads to the
residuals which are dominated by the random fluctuations only. An expression
for the mean-square residuals after subtracting the best-fitting solution
for the estimates (\ref{111}) is given by the formula \vspace{0.3 cm} 
\begin{equation}
<r^{2}({\Delta T})>=\frac{1}{mN}{\displaystyle \sum_{i=1}^{mN}}%
{\displaystyle \sum_{j=1}^{mN}}K(t_{i},t_{j})R(t_{i},t_{j}),\vspace{0.3 cm}  
\label{112a}
\end{equation}\vspace{0.3 cm}
where the function 
\begin{equation}\vspace{0.3 cm}
K(t_{i},t_{j})=\delta _{ij}-
{\displaystyle \sum_{a=1}^{14}}\Xi_{a}(t_{i})\psi _{a}(t_{j}),\vspace{0.3 cm}  \label{112b}
\end{equation}\vspace{0.3 cm}
is called the filter function (Blandford {\it et al.} 1984). 
We have proved (Kopeikin 1999b) that the post-fit residuals depend only on the 
stationary part of the noise (even in the case when the non-stationary part of the autocovariance function is present) which means that Eq. (\ref{112a})  can be written down as 
\begin{equation}\vspace{0.3 cm}
<r^{2}({\Delta T})>=
-\frac{1}{mN}{\displaystyle \sum_{a=1}^{14}}{\displaystyle 
\sum_{b=1}^{14}}L_{ab}^{-1}\left[
{\displaystyle \sum_{i=1}^{mN}}{\displaystyle \sum_{j=1}^{mN}}
\psi _{a}(t_{i})\psi
_{b}(t_{j})R^{-}(t_{i},t_{j})\right] 
. \label{vlbi}
\end{equation}\vspace{0.3 cm}
For this reason, methods of spectral analysis in frequency domain can be 
applied for analyzing residuals without any restrictions. Let us note that the
explicit dependence of TOA residuals on the total span of observations is contained
in (Kopeikin 1999b).
      
\section{Fourier Transform of Fitting Functions}

We define the Fourier transform of the fitting functions $\psi_a(t)$ as
\vspace{0.3 cm}
\begin{equation}
\label{summa}
\tilde{\Psi}_a(f,m,N)=\displaystyle{\sum_{j=1}^{mN}}\psi_a(t_j)
\exp(-2\pi i f t_j),
\end{equation}
\vspace{0.3 cm}
where $f$ is the Fourier frequency measured in units inversly
proportional to the units of measurement of time $t$. We measure time in units 
of the orbital phase $u=n_b t$, that is in
radians. Then the frequency $\omega =2\pi f$ is dimensionless and is 
measured in units of the orbital frequency $n_b$. One notes the Fourier transform
of the fitting functions depends on three arguments: the Fourier frequency $f$, the total number of 
orbital revolutions $N$, and the rate of observations $m$.   

When the total number of observational points, $mN$, is large enough we can 
approximate the sum in Eq. (\ref{summa}) by the integral (Kopeikin 1999b)
\vspace{0.3 cm}
\begin{equation}
\label{integ}
\tilde{\Psi}_a(\omega,m,N)=\frac{m}{2\pi}\tilde{\psi}_a(\omega,N),
\end{equation}
\vspace{0.3 cm}
\begin{equation}
\label{in}
\tilde{\psi}_a(\omega,N)=
\displaystyle{\int_{-\pi N}^{\pi N}}\psi_a(u) \exp(-i \omega u) du.
\label{ft1}
\end{equation}
We note that $\tilde{\psi}_a(-\omega)=
\tilde{\psi}_a^{\ast}(\omega)$, where the asterisk denotes a complex 
conjugation.
Replacing the sum over observational points by the integral with respect to
time (the orbital phase) makes our calculations equivalent to the case of continuous observations with uniform distribution over time.

The following formulae are also of use in practical computations
\vspace{0.3 cm}
\begin{equation}
\hspace{0.3 cm}\tilde{\psi}_a(\omega,N)=\left\{\begin{array}{ll}
2\displaystyle{\int_{0}^{\pi N}}\psi_a(u)
\cos(\omega u) du,\hspace{0.3 cm} \mbox{if index $a=1,3,5,...$}\nonumber\\ \\ \nonumber
-2i\displaystyle{\int_{0}^{\pi N}}\psi_a(u)
\sin(\omega u) du,\hspace{0.3 cm}\mbox{if index $a=2,4,6,...$.}
\end{array}\right.
\label{ft3}
\end{equation}
\vspace{0.3 cm}
These expressions shows that the fitting functions with odd indices are real
and those with even ones are complex. 

Let us denote ${\rm T}=\pi N$ and $z=\omega {\rm T}$. One notices that ${\rm T}$ relates to the total span of observation using equation ${\rm T}=n_b\Delta T/2$.
The Fourier transform of fitting functions takes the form
\begin{eqnarray}
\tilde{\psi}_1(\omega)&=&2{\rm T}\tilde{\phi}_1(z),
\label{f1}
\end{eqnarray}
\begin{eqnarray}
\tilde{\psi}_2(\omega)&=&2i{\rm T}^2\tilde{\phi}_2(z),
\label{f2}
\end{eqnarray}
\begin{eqnarray}
\tilde{\psi}_3(\omega)&=&2{\rm T}^3\tilde{\phi}_3(z),
\label{f3}
\end{eqnarray}
\begin{eqnarray}
\tilde{\psi}_4(\omega)&=&2i{\rm T}^4\tilde{\phi}_4(z),
\label{f4}
\end{eqnarray}
\begin{eqnarray}
\tilde{\psi}_5(\omega)&=&2{\rm T}^5\tilde{\phi}_5(z),
\label{f5}
\end{eqnarray}
\begin{eqnarray}
\tilde{\psi}_6(\omega)&=&2i{\rm T}^6\tilde{\phi}_6(z),
\label{f6}
\end{eqnarray}
\begin{eqnarray}
\tilde{\psi}_7(\omega)&=&{\rm T}\left[\tilde{\phi}_1(z+{\rm T})+
\tilde{\phi}_1(z-{\rm T})\right],
\label{f7}
\end{eqnarray}
\begin{eqnarray}
\tilde{\psi}_8(\omega)&=&i{\rm T}\left[\tilde{\phi}_1(z+{\rm T})-
\tilde{\phi}_1(z-{\rm T})\right],
\label{f8}
\end{eqnarray}
\begin{eqnarray}
\tilde{\psi}_9(\omega)&=&{\rm T}^2\left[\tilde{\phi}_2(z-{\rm T})
-\tilde{\phi}_2(z+{\rm T})\right],
\label{f9}
\end{eqnarray}
\begin{eqnarray}
\tilde{\psi}_{10}(\omega)&=&i{\rm T}^2\left[\tilde{\phi}_2(z-{\rm T})
+\tilde{\phi}_2(z+{\rm T})\right],
\label{f10}
\end{eqnarray}
\begin{eqnarray}
\tilde{\psi}_{11}(\omega)&=&{\rm T}^3\left[\tilde{\phi}_3(z+{\rm T})+
\tilde{\phi}_3(z-{\rm T})\right],
\label{f11}
\end{eqnarray}
\begin{eqnarray}
\tilde{\psi}_{12}(\omega)&=&i{\rm T}^3\left[\tilde{\phi}_3(z+{\rm T})-
\tilde{\phi}_3(z-{\rm T})\right],
\label{f12}
\end{eqnarray}
\begin{eqnarray}
\tilde{\psi}_{13}(\omega)&=&{\rm T}^4\left[\tilde{\phi}_4(z-{\rm T})-
\tilde{\phi}_4(z+{\rm T})\right],
\label{f13}
\end{eqnarray}
\begin{eqnarray}
\tilde{\psi}_{14}(\omega)&=&i{\rm T}^4\left[\tilde{\phi}_4(z-{\rm T})+
\tilde{\phi}_4(z+{\rm T})\right].
\label{f14}
\end{eqnarray}
\vspace{0.5 cm}
where functions $\phi_a(z)$ ($a=1,2,...,6)$ have the following form
\begin{eqnarray}
\tilde{\phi}_1(z)&=&\frac{\sin z}{z},
\label{ff1}
\end{eqnarray}

\begin{eqnarray}
\tilde{\phi}_2(z)&=&
\frac{\cos z }{z}-
\frac{\sin z}{z^2},
\label{ff2}
\end{eqnarray}

\begin{eqnarray}
\tilde{\phi}_3(z)&=&\frac{\sin z}{z}+
\frac{2 \cos z }{z^2}-
\frac{2 \sin z }{z^3},
\label{ff3}
\end{eqnarray}

\begin{eqnarray}
\tilde{\phi}_4(z)&=&\frac{\cos z }{z}-
\frac{3 \sin z }{z^2}-
\frac{6\cos z}{z^3}+
\frac{6 \sin z}{z^4},
\label{ff4}
\end{eqnarray}

\begin{eqnarray}
\tilde{\phi}_5(z)&=&\frac{\sin z}{z}+
\frac{4\cos z}{z^2}-
\frac{12\sin z}{z^3}-
\frac{24\cos z }{z^4}+
\frac{24\sin z }{z^5},
\label{ff5}
\end{eqnarray}

\begin{eqnarray}
\tilde{\phi}_6(z)&=&\frac{\cos z}{z}-
\frac{5\sin z }{z^2}-
\frac{20\cos z }{z^3}+
\frac{60\sin z}{z^4}+
\frac{120\cos z }{z^5}-
\frac{120\sin z}{z^6}.
\label{ff6}
\end{eqnarray}

Simple inspection reveals that these functions are linear combinations of the spherical Bessel
functions $j_a(z)$ defined as (Korn \& Korn 1968, section 21.8-8)
\begin{equation}\label{bessel}
j_a(z)=z^a\left(-\frac{1}{z}\frac{d}{dz}\right)^a \frac{\sin
z}{z},\quad (a=0,1,2,...)\;.
\end{equation}
The Bessel functions have a polynomial
behaviour \footnote{The function
$j_a(z)=\frac{z^a}{(2a+1)!!}+O(z^{a+2})$ for $z\ll 1$ ($a=0,1,2,...$).} near the point $z=0$, and 
then oscillate with monotonically
decreasing amplitude. Asymptotic expansion of the sperical Bessel functions for
large values of the variable $z$ (that is directly proportional to the Fourier frequency) are
given by the formula 
\begin{equation}
j_a(z)\approx \frac{\sin\left(z-\displaystyle{\frac{\pi a}{2}}\right)}{z}.
\end{equation} 
It is worth emphasizing that
the maximum value for any of these functions can not be larger than 1.

The fitting functions $\tilde\phi_a(z)$, 
expressed in terms of the spherical Bessel
functions, assume the form
\begin{eqnarray}
\tilde\phi_1(z)&=&j_0(z),\\\nonumber\\
\tilde\phi_2(z)&=&-j_1(z),\\\nonumber\\
\tilde\phi_3(z)&=&\frac{1}{3}j_0(z)-\frac{2}{3}j_2(z),\\\nonumber\\
\tilde\phi_4(z)&=&-\frac{3}{5}j_1(z)+\frac{2}{5}j_3(z),\\\nonumber\\
\tilde\phi_5(z)&=&\frac{1}{5}j_0(z)-\frac{4}{7}j_2(z)+\frac{8}{35}j_4(z),\\\nonumber\\
\tilde\phi_6(z)&=&-\frac{3}{7}j_1(z)+\frac{4}{9}j_3(z)-\frac{8}{63}j_5(z).
\end{eqnarray}

\section{Fourier Transform of the Covariance Matrix} 

In order to calculate the covariance matrix we need to know the 
Fourier transform of the dual functions 
$\Xi_a(t)$. The transform is defined in accordance with definition (\ref{ert}) 
of the dual functions and takes the form 
\vspace{0.3 cm}
\begin{equation}
\tilde{\Xi}_a(f,m,N)=\displaystyle{
\sum_{c=1}^{14}}L_{ac}^{-1}\tilde{\Psi}_c(f,m,N),
\label{xi}
\end{equation}
and the cross-orthonormal condition in the frequency domain is given by the
integral
\vspace{0.3 cm}
\begin{equation}
\displaystyle{\int_{0}^{\infty}}\tilde{\Xi}_a(f,m,N)\tilde{\Psi}_b(f,m,N)df=
\frac{1}{2}\delta_{ab}.
\label{dual}
\end{equation}

In the limit of continuous observations it is convenient to 
introduce the matrix $C_{ab}=\frac{2\pi}{m}L_{ab}$ instead of the information matrix $L_{ab}$.
The explicit expression for the matrix $C_{ab}$ is given by the integral:
\vspace{0.3 cm}
\begin{equation}
C_{ab}=\displaystyle{\int_{-\pi N}^{\pi N}}\psi_a(u) \psi_b(u) du,
\end{equation}
and the result of evaluation of this integral is given in the paper by Kopeikin
(1999b) in Tables 5 and 6.
Then we have  $L_{ab}^{-1}=(2\pi/m)C_{ab}^{-1}$, and the Fourier transform of the dual function
$\tilde{\Xi}(f)$ can be recast as
\begin{equation}
\tilde{\Xi}_a(f,N)=\displaystyle{
\sum_{b=1}^{14}}C_{ab}^{-1}\tilde{\psi}_b(f,N).
\label{gfl}
\end{equation}
\vspace{0.3 cm}
Hence, comparing the eq. (\ref{gfl}) with (\ref{xi}) one concludes that 
in the limit of continuous observations the Fourier transform of the 
dual functions depends only on the
Fourier frequency and the total number of orbital revolutions as was expected.
It is more insightful to express the dual functions (\ref{gfl}) in terms of 
the spherical Bessel functions (\ref{bessel}). The expressions obtained are
rather unwieldy, and, for this reason they are given in Appendix A.  

Making use of the definition of the Fourier transforms of the stationary part of the 
autocovariance function (\ref{2}) and the dual functions (\ref{xi})
we obtain the stationary part of the covariance matrix
$M_{ab}^{-}(m,N)$ (see Eq. (\ref{112}) for its definition) expressed as follows
\begin{equation}
M_{ab}^{-}=\displaystyle{\int_{\varepsilon}^{\infty}}S(f)H_{ab}(f,m,N)df,
\label{mab}
\end{equation}
where $H_{ab}(f,m,N)$ is the transfer function given by the expression 
\begin{equation}
H_{ab}(f,m,N)=
\tilde{\Xi}_a(f,m,N)\tilde{\Xi}_b^{\ast}(f,m,N)+
\tilde{\Xi}_a^{\ast}(f,m,N)\tilde{\Xi}_b(f,m,N),
\label{h}
\end{equation}
and an asterisk denotes complex conjugation. For numerical computations of 
$M_{ab}^{-}$ the following formula can be used in practical computations
\vspace{0.3 cm}
\begin{eqnarray}\label{cvb}
M_{ab}^{-}(m,N)&=&\frac{h_n}{(2\pi)^n}\int_{\varepsilon}^{\Lambda}H_{ab}
(f,m,N)f^{-n}df+\\\nonumber\\\nonumber&&
\frac{1}{2}h_n \left[B_0(\varepsilon)H_{ab}(\varepsilon,m,N)+
\varepsilon^2 B_2(\varepsilon) H_{ab}^{(2)}(\varepsilon,m,N)+
\varepsilon^4 B_4(\varepsilon) H_{ab}^{(4)}(\varepsilon,m,N)+...\right],
\end{eqnarray}
where derivatives of $H_{ab}$ are taken with respect to the Fourier frequency, 
ellipses denote terms of negligible influence on the result of the
computation, and $\Lambda$ is the upper cutoff frequency arising from the
sampling theorem and is inversly proportional to the minimal time between
subsequent observational sessions.  We emphasize that the matrix $M_{ab}^{-}(m,N)$ does not depend on the lower cutoff frequency because all contributions from the lower limit of the integral in Eq. (\ref{cvb}) are canceled out by corresponding terms from the series.

\section{Fourier Transform of the Filter Function}

Timing 
residuals are expressed in terms of the Fourier transform of the filter function.  Employing eqs. (\ref{2}),
(\ref{vlbi}), and (\ref{ft1}) yields for the residuals\vspace{0.3 cm}
\begin{equation}
<r^2> =
2\displaystyle{\int_{\varepsilon}^{\infty}}S(f)K(f,m,N)df,
\label{resid}
\end{equation}
\vspace{0.3 cm}
where is the Fourier transform of the filter function (\ref{112b})
\begin{equation}
K(f,m,N)=1-\frac{1}{2mN}\displaystyle{\sum_a^{14}}
\left[\tilde{\Xi}_a(f,m,N)\tilde{\Psi}_a^{\ast}(f,m,N)+
\tilde{\Xi}_a^{\ast}(f,m,N)\tilde{\Psi}_a(f,m,N)\right],
\label{filt}
\end{equation}
is the Fourier transform of the filter function defined in Eq. (\ref{112b}).
In the limit of continuous observations there is no dependence on the frequency
of observations, $m$, so that one obtains
\begin{equation}
K(f,N)=1-\frac{1}{4{\rm T}}\displaystyle{\sum_a^{14}}
\left[\tilde{\Xi}_a(f,N)\tilde{\psi}_a^{\ast}(f,N)+
\tilde{\Xi}_a^{\ast}(f,N)\tilde{\psi}_a(f,N)\right].
\label{filter}
\end{equation}

Plots of the Fourier transform (\ref{filter}) of the filter function 
$K(f)$ are shown in Appendix D for different numbers of orbital revolutions 
$N$. In any case the filter function is approximately equal to 1 until the
frequency is higher than $1/(\Delta T)$, and rapidly decreases in amplitude 
as the frequency approaches zero. We also note that the plot of the 
Fourier transform of the filter function clearly shows the additional dip
near the orbital frequency. The dips near zero and orbital frequencies  
get narrower as the number of observational points increases. 

\section{Spectral Sensitivity of Timing Observations of 
Millisecond and Binary Pulsars}

Analytical expressions and graphical representations of Fourier transforms of
dual functions and timing residuals help us to understand in more detail
the spectral sensitivity of single and binary pulsars to different frequency bands
in the spectral decomposition of the noise.
First, let us consider behaviour of the Fourier transform of fitting functions
near zero and the orbital frequencies.

It is easy to confirm after making use of Taylor expansion of exponential
function in (\ref{ft1}) near $\omega\simeq 0$ that\vspace{0.3 cm}
\begin{equation}
\tilde{\psi}_a(\omega)= \left\{\begin{array}{ll}
C_{a1}-\frac{1}{2}\omega^2 C_{a3}+\frac{1}{24}\omega^4 C_{a5}+\omega^6
p_{a},&\mbox{if a=1,3,5,...}\\ \\
i\left(
-\omega C_{a2}+\frac{1}{6}\omega^3 C_{a4}-\frac{1}{120}\omega^5 C_{a6}+\omega^7
p_{a}\right),& \mbox{if a=2,4,6,...,}
\end{array}\right.
\label{expanzero}
\end{equation}\vspace{0.3 cm}
where $p_a$ is a residual term depending only on the total number of orbital
revolutions, $N$.
A Taylor expansion of the Fourier transform of the fitting functions near the orbital
frequency yields\vspace{0.3 cm}
\begin{equation}
\tilde{\psi}_a(\omega)= \left\{\begin{array}{ll}
C_{a7}+(\omega-1) C_{a9}-\frac{(\omega-1)^2}{2} C_{a.11}-
\frac{(\omega-1)^3}{6} C_{a.13}+(\omega-1)^4 
q_{a},&\mbox{if a=1,3,...}\\ \\
i\left[
-C_{a8}+(\omega-1) C_{a.10}+\frac{(\omega-1)^2}{2} C_{a.12}-
\frac{(\omega-1)^3}{6} C_{a.14}+(\omega-1)^4 q_{a}\right],& \mbox{if a=2,4,...},
\end{array}\right.
\label{expanorb}
\end{equation}
where $q_a$ is a residual term depending only on the total number of orbital
revolutions, $N$ (recall that the frequency is measured in units 
of the orbital frequency $n_b$). Applying to Eqs.
(\ref{expanzero})--(\ref{expanorb}) the definition of the dual functions
(\ref{gfl}) in the limit of continuos observations yields the asymptotic behaviour
of the dual functions
\vspace{0.3 cm}
\begin{equation}
\tilde{\Xi}_a(\omega)= \left\{\begin{array}{ll}
\delta_{a1}-\frac{1}{2}\omega^2 \delta_{a3}+\frac{1}{24}\omega^4 \delta_{a5}+
\omega^6 P_{a},&\mbox{if a=1,3,5,...}\\ \\
i\left(
-\omega \delta_{a2}+\frac{1}{6}\omega^3 \delta_{a4}-\frac{1}{120}\omega^5
\delta_{a6}+\omega^7 P_{a}\right)
,& \mbox{if a=2,4,6,...,}
\end{array}\right.
\label{dfzero}
\end{equation}\vspace{0.3 cm}
near zero frequency, and \vspace{0.3 cm}
\begin{equation}
\tilde{\Xi}_a(\omega)= \left\{\begin{array}{ll}
\delta_{a7}+(\omega-1) \delta_{a9}-\frac{(\omega-1)^2}{2} C_{a.11}-
\frac{(\omega-1)^3}{6} 
\delta_{a.13}+(\omega-1)^4 Q_{a},&\mbox{if a=1,3,...}\\ \\
i\left[-\delta_{a8}+(\omega-1) \delta_{a.10}+\frac{(\omega-1)^2}{2} 
\delta_{a.12}-
\frac{(\omega-1)^3}{6} \delta_{a.14}+(\omega-1)^4 Q_{a}\right],& 
\mbox{if a=2,4,...},
\end{array}\right.
\label{dforb}
\end{equation}
near the orbital frequency, where $P_a$ and $Q_a$ are residual terms. 
Table 2 shows the asymptotic behaviour of the residual terms of the dual 
functions.

Now we can study the asymptotic behaviour of the filter function $K(f)$ as defined by eq.
(\ref{filter}). Taking into account the fact that
$\sum_{b=1}^{14} C_{ab}^{-1}C_{bc}=\delta_{ac}$ where $\delta_{ac}$ is the unit matrix, we
obtain\vspace{0.3 cm}
\begin{equation}
K(f) = \left\{\begin{array}{ll}
{\rm b_0} \cdot \omega^6, &\mbox{when $\omega \rightarrow 0$}\\ \\
{\rm b_1} \cdot (\omega-1)^4, &\mbox{when $\omega \rightarrow 1$}
\end{array}\right.
\label{tfil}
\end{equation}\vspace{0.3 cm}
where ${\rm b}_0$ and ${\rm b}_1$ are numerical constants which can be calculated precisely but are not important for the discussion present in this section.
Such a dependence of the filter function $K(f)$ on the frequency $f$ significantly reduces the amount of the
detected red-noise power below the cutoff frequency $f_{\rm cutoff} \simeq \alpha_c {\rm
T}^{-1}$ and in the frequency band $1-\alpha_b {\rm T}^{-1} \leq f 
\leq 1+\alpha_b {\rm T}^{-1}$ lying near the orbital frequency. Here constant
coefficents $\alpha_c$ and $\alpha_b$ can be determined by comparing
the calculations of the mean value of the timing residuals in time and frequency
domains (Kopeikin 1997a). The low Fourier frequencies
in the noise power spectrum are fitted away by the polynomial fit for the spin-down
parameters of the observed pulsar. The Fourier frequencies close to the orbital frequency
are fitted away by the fit for the orbital parameters of the pulsar. The amount of
noise power remaining in timing residuals 
after completion of fitting procedure~\footnote{It is useful to compare calculations of the residuals given in the present paper with those given in (Kopeikin 1999b.} is estimated by the expressions \vspace{0.3
cm}
\begin{equation}
< r^2 > = 
2\displaystyle{\int_{\frac{\alpha_c}{\rm T}}^
{1-\frac{\alpha_b}{\rm T}}}S(f)df+
2\displaystyle{\int_{1+\frac{\alpha_b}{\rm T}}^{\infty}}S(f)df =
\frac{2h_n}{(2 \pi)^n}\left(\frac{\alpha_c^{1-n}{\rm T}^{n-1}}{n-1}-
\frac{2\alpha_b}{{\rm T}}\right)+O\left(\frac{1}{{\rm T}^3}\right)\;,\quad(n>1)
\label{uuu}
\end{equation}and
\begin{equation}
< r^2 > =2\displaystyle{\int_{\frac{\alpha_c}{\rm T}}^
{1-\frac{\alpha_b}{\rm T}}}S(f)df+
2\displaystyle{\int_{1+\frac{\alpha_b}{\rm T}}^{\infty}}S(f)df =\frac{h_1}{\pi}\left(\ln{\rm T}-\frac{2\alpha_b}{{\rm T}}\right)+O\left(\frac{1}{{\rm T}^3}\right)\;,\quad(n=1)
\label{dopf}
\end{equation}
\vspace{0.3 cm}
where for the calculation of the first and the second integrals we have taken the very first terms of the spectra~\footnote{Delta-functions and their derivatives do not contribute to the integrals in the approximation under consideration.} given in Eqs. (\ref{7})--(\ref{rww})   and assumed that $\alpha_b/{\rm T}\ll 1$ and $\alpha_b/{\rm T}\ll 1$ so that these quantities are used as small parameters of the Taylor expansion of the integrals under calculation. We draw attention of the reader that in real signal processing the infinite-frequency limit in Eqs.(\ref{uuu}), (\ref{dopf}) is, in fact, inversely proportional to the sampling time of observations.

The second term in the right hand side of 
Eqs. (\ref{uuu}), (\ref{dopf}) shows amount of noise absorbed by fitting
orbital parameters. It is negligibly small compared to the first term in the
right hand side and
can be neglected in practice. Hence, we conclude that the post-fit
timing residuals can be used for the estimation of the amount of red noise and its
spectrum in frequency band just from $\alpha_c {\rm T}^{-1}$ up to infinity,
irrespectively of whether the pulsar is binary or not. This reasoning puts on
firm ground the estimates of spectral sensitivity of timing observations
and the cosmological parameter $\Omega_g$, characterizing energy density of
stochastic gravitational waves in early universe, made by Kaspi {\it et al.}
(1994) and Camilo {\it et al.} (1994) using observations of binary pulsars PSR
B1855+09 and PSR J1713+0747 respectively. 

Analysis of the spectral sensitivity of the estimates of variances of spin-down and
orbital parameters is more cumbersome.
We are interested in which frequencies give the biggest contribution to the
variances. This is important to know, for example, if we want to use
variances of certain orbital parameters for setting the fundamental upper 
limit on $\Omega_g$ in the ultra-low frequency domain  (Kopeikin 1997a). 
Analitical calculations reveal the leading terms in the asymptotic 
expansions of the dual functions near zero frequencies which are present in Table 2.    
\begin{table*}
\begin{minipage}{130mm}
\centering
\caption{Asymptotic behaviour of residual terms of 
the dual functions $\tilde{\Xi}_a$ near zero
and orbital frequencies. Constant ${\rm h}=\cos{\rm T}=(-1)^N$.}
\label{tab:11}
\vspace{3 mm}
\begin{tabular}{|ccc|}
\hline \\ \\
Dual function &Residual term $P_a$&Residual term $Q_a$ \\ \hline
&  \\ 
$\tilde{\Xi}_{1}$&$-\frac{1}{33264} {\rm T}^6$&$
\frac{19}{56}
{\rm T}^2{\rm h}$ \\ 
&  \\ 
$\tilde{\Xi}_{2}$&$\frac{1}{61776} {\rm T}^6$&
$-\frac{1}{8}{\rm T}^2 {\rm h}$\\ 
&  \\ 
$\tilde{\Xi}_{3}$&$\frac{1}{1584}{\rm T}^4$&
$-\frac{17}{4}{\rm h}$\\ 
&  \\ 
$\tilde{\Xi}_{4}$&$-\frac{1}{6864} {\rm T}^4$&
$-\frac{3}{4} {\rm h}$\\ 
&  \\ 
$\tilde{\Xi}_{5}$&$-\frac{1}{528}{\rm T}^2$&
$-\frac{45}{8}\frac{{\rm h}}{{\rm T}^2}$\\ 
&  \\  
$\tilde{\Xi}_{6}$&$\frac{1}{3120}{\rm T}^2$&
$\frac{33}{40}\frac{{\rm h}}{{\rm T}^2} $\\ 
&  \\
$\tilde{\Xi}_{7}$ &$-\frac{1}{165}{\rm T}^4{\rm h}$&
$-\frac{1}{280} {\rm T}^4$ \\ 
&  \\ 
$\tilde{\Xi}_{8}$ &$\frac{1}{45045}{\rm T}^6{\rm h}$&
$\frac{1}{280} {\rm T}^4$ \\ 
&  \\ 
$\tilde{\Xi}_{9}$ &$-\frac{1}{693}{\rm T}^4{\rm h}$&
$-\frac{1}{28}{\rm T}^2$ \\ 
&  \\  
$\tilde{\Xi}_{10}$ &$-\frac{1}{273}{\rm T}^4{\rm h}$&
$\frac{1}{28} {\rm T}^2$ \\ 
&  \\ 
$\tilde{\Xi}_{11}$ &$-\frac{19}{693}{\rm T}^2{\rm h}$&
$\frac{1}{28} {\rm T}^2$ \\ 
&  \\
$\tilde{\Xi}_{12}$ &$\frac{1}{9009}{\rm T}^4{\rm h}$&
$-\frac{1}{28} {\rm T}^2$ \\ 
&  \\  
$\tilde{\Xi}_{13}$ &$\frac{1}{297}{\rm T}^2{\rm h}$&
$\frac{1}{12}$ \\ 
&  \\
$\tilde{\Xi}_{14}$ &$\frac{31}{3861}{\rm T}^2{\rm h}$&
$-\frac{1}{12}$ \\ 
&  \\ \\ \\
\hline  
\end{tabular}
\end{minipage}
\end{table*}
The squares of the dual functions $\tilde{\xi}_a$ appearing 
in eqs. (\ref{dz0})-(\ref{dz14}) and eqs. (\ref{dzT})-(\ref{dzTT}) are rather
complicated. Their behaviour is periodic with bumps both near zero and the orbital 
frequencies and with occilating behaviour which amplitude is rapidly decaying far outside of these 
frequencies. We determined that the amount of noise absorbed by fitting of the spin-down parameters 
near zero frequency 
is essentially bigger than that near the orbital one. On the other hand, the amount of noise absorbed by fitting of the orbital parameters is substantial for the Fourier freqiencies lying near the orbital frequency. Thus, there are two spectral windows in which parameter' variances of the timing model are most sensitive
to the stochastic red noise and they are located near zero and the orbital frequency. These windows are restricted by two
frequency intervals, (0,$\frac{\alpha}{\rm T}$), and, 
($1-\frac{\alpha_{-}}{\rm T},1+\frac{\alpha_{+}}{\rm T}$), respectively, where 
constant
coefficients $\alpha$, $\alpha_{-}$, $\alpha_{+}$ can be calculated by  
comparing calculations
of variances of the fitting parameters in time and frequency domains. The variances of the fitting parameters
depend only on the total span, $\Delta T$, of observations (Kopeikin 1997b). 
Comparing the dependence of the variances on $\Delta T$ calculated in time domain with 
that calculated in the frequency domain with the integrals truncated by the two frequency windows, reveals the relative importance in contribution of different frequencies to the integrated values of the variances of the parameters given by 
two integrals
\vspace{0.3 cm}
\begin{equation}
{\rm I_1} \sim h_n \displaystyle{\int_{\varepsilon}^{\frac{\alpha}{\rm T}}}
|\tilde{\Xi}_a(f)|^2 \left[\displaystyle{
\frac{1}{(2\pi f)^n}}+\displaystyle{\sum_{k=0}^\infty}B_{2k}(\varepsilon) 
\varepsilon^{2k}\delta^{(2k)}(f-\varepsilon)\right] df,
\label{i1}
\vspace{0.3 cm}
\end{equation}
and\vspace{0.3 cm}
\begin{equation}
{\rm I_2} \sim \frac{h_n}{(2\pi)^n}
\displaystyle{\int_{1-\frac{\alpha_{-}}{\rm T}}^
{1+\frac{\alpha_{+}}{\rm T}}}
|\tilde{\Xi}_a(f)|^2 f^{-n} df,
\label{i2}
\vspace{0.3 cm}
\end{equation} 

The first integral describes the contribution of low frequencies to the parameter's
variances. The second integral gives the contribution to the parameter's variances from 
the frequencies lying near the orbital frequency. It is worth emphasizing that the terms 
with delta functions and their derivatives in Eq. (\ref{i1}) 
cancel out all terms depending on the cutoff frequency $\varepsilon$ and 
diverging as $\varepsilon$ goes to zero. 
This cancellation has been expected since we modified the spectrum of
red noise to avoid the appearance of all divergent terms which have no 
physical meaning. We don't give here the results of the calculation of numerical values
of constants $\alpha$, $\alpha_{-}$, $\alpha_{+}$ because they are not so
important for general conclusions. The asymptotic behaviour of the dual functions
near zero and the orbital frequency is enough informative to see which frequncy band is the
most important for giving contributions to the corresponding integrals and the 
parameter's variances. 
The time dependence of two integrals is shown in Table \ref{tab12}
up to a not so important constant.   
\begin{table*}
\begin{minipage}{130mm}
\centering
\caption{Comparative contribution of different frequency bands to variances of
spin-down ($a=1,2,...,6$) and orbital ($a=7,8,...,14$) parameters $\beta_a$.
Number $n=1,2,...,6$ denotes the spectral index of corresponding red noise.
Time dependence of all variances completely coincides with that which was
obtained by calculations in time domain as given in (Kopeikin 1999b).}
\label{tab12}
\vspace{3 mm}
\begin{tabular}{|ccc|}
\hline \\ 
Variance of parameter&Contribution of integral ${\rm I_1}$&
Contribution of integral ${\rm I_2}$ \\ \hline
&  \\ 
$\sigma_{\beta_1}^2$&$\sim {\rm T}^{n-1}$&$\sim {\rm T}^{-3}$ \\ 
&  \\ 
$\sigma_{\beta_2}^2$&$\sim {\rm T}^{n-3}$&$\sim {\rm T}^{-5}$ \\ 
&  \\ 
$\sigma_{\beta_3}^2$&$\sim {\rm T}^{n-5}$&$\sim {\rm T}^{-7}$ \\ 
&  \\ 
$\sigma_{\beta_4}^2$&$\sim {\rm T}^{n-7}$&$\sim {\rm T}^{-9}$ \\ 
&  \\ 
$\sigma_{\beta_5}^2$&$\sim {\rm T}^{n-9}$&$\sim {\rm T}^{-11} $\\ 
&  \\ 
$\sigma_{\beta_6}^2$&$\sim {\rm T}^{n-11}$&$\sim {\rm T}^{-13}$ \\ 
&  \\
$\sigma_{\beta_7}^2$&$\sim {\rm T}^{n-5}$&$\sim {\rm T}^{-1}$ \\ 
&  \\ 
$\sigma_{\beta_8}^2$&$\sim {\rm T}^{n-3}$&$\sim {\rm T}^{-1}$ \\ 
&  \\ 
$\sigma_{\beta_9}^2$&$\sim {\rm T}^{n-5}$&$\sim {\rm T}^{-3}$ \\ 
&  \\  
$\sigma_{\beta_{10}}^2$&$\sim {\rm T}^{n-7}$&$\sim {\rm T}^{-3}$ \\ 
&  \\ 
$\sigma_{\beta_{11}}^2$&$\sim {\rm T}^{n-9}$&$\sim {\rm T}^{-5}$ \\ 
&  \\
$\sigma_{\beta_{12}}^2$&$\sim {\rm T}^{n-7}$&$\sim {\rm T}^{-5}$ \\ 
&  \\  
$\sigma_{\beta_{13}}^2$&$\sim {\rm T}^{n-9}$&$\sim {\rm T}^{-7}$ \\ 
&  \\
$\sigma_{\beta_{14}}^2$&$\sim {\rm T}^{n-11}$&$\sim {\rm T}^{-7}$ \\ 
&  \\ \\
\hline  
\end{tabular}
\end{minipage}
\end{table*}
The behavior of the variances of the first three spin-down parameters is not
physically interesting because they are contaminated and strongly biased by the presence of the non-stationary 
part of the red noise (Kopeikin 1997b). For the rest of the spin-down parameters one 
observes that the contribution of the noise energy from  low frequencies to the variances 
of the parameters is dominating. However, the situation is not so simple in 
the case of the orbital parameters. 
One can see that in the case of red noise having spectral index $n \leq 4$,
the contribution of the noise energy from the orbital frequency interval 
($1-\frac{\alpha_{-}}{\rm T},1+\frac{\alpha_{+}}{\rm T}$) can be equal to or
even bigger than that from the low frequency band. Only when the spectral index 
of noise is $n\geq 5$ does the contribution of the noise energy of low frequencies to the 
variances of orbital parameters begins to dominate. 

It is worth noting that the
timing noise with the spectral index $n=5$ is produced by the cosmological
gravitational wave background. The fact that for this noise 
low-frequencies give the main contribution to the variances of orbital parameters
confirms our early statement (Kopeikin 1997a) that the measurement of the variances of
orbital parameters showing secular evolution probes the ultra-low frequency band
of the cosmological gravitational wave background. Hence, these variances can be 
used for setting an upper limit on the cosmological parameter $\Omega_g$ in this
frequency range, in contrast to timing residuals which test only the low-frequency 
band of the background noise  as explained in
(Kopeikin 1997a).  
\section{Acknowledgments}
 
We are grateful to N. Wex
for numerous fruitful discussions which have helped to improve the 
presentation of this manuscript. 
We thank H. Lambert and A. Corman
for careful reading of the manuscript and valuable comments. 
We are indebted to the anonymous referee for pointing out new references and for suggestions which helped to shorten and clarify the article.

\appendix

\section{Explicit expressions for the dual functions}

In this appendix we give explicit expressions for the dual functions. Using
definition (\ref{gfl}) and elements of inverse matrix $C_{ab}^{-1}$ from the
paper (Kopeikin 1999b, Tables 5 and 6) one obtains
\vspace{0.3 cm}
\begin{eqnarray}\label{dfu1}
\tilde{\Xi}_1(z)&=&j_0(z)+\frac{5}{2}j_2(z)+\frac{27}{8}j_4(z)+
\\\nonumber\\\nonumber\mbox{}&&
+\frac{15{\rm h}}{8\rm T}\biggl\{3\left[j_1(z+{\rm T})-j_1(z-{\rm T})\right]
-7\left[j_3(z+{\rm T})-j_3(z-{\rm T})\right]\biggr\}\\\nonumber\\\nonumber\mbox{}&&-
\frac{45{\rm h}}{8{\rm T}^2}\biggl\{3\left[j_0(z+{\rm T})+ j_0(z-{\rm T})\right]
-20\left[j_2(z+{\rm T})+j_2(z-{\rm T})\right]\biggr\}\;,
\end{eqnarray}

\begin{eqnarray}\label{dfu2}
\frac{1}{i}\tilde{\Xi}_2(z)&=&-\frac{3}{\rm T}\left[
j_1(z)+\frac{7}{2}j_3(z)+\frac{55}{8}j_5(z)\right]+
\\\nonumber\\\nonumber\mbox{}&&
\frac{105{\rm h}}{8{\rm T}^2}
\biggl\{j_0(z+{\rm T})-j_0(z-{\rm T})
-5\left[j_2(z+{\rm T})-j_2(z-{\rm T})\right]\biggr\}+\\\nonumber\\\nonumber\mbox{}&&
\frac{105{\rm h}}{8{\rm T}^3}\biggl\{51\left[
j_1(z+{\rm T})+ j_1(z-{\rm T})\right]
-154\left[j_3(z+{\rm T})+j_3(z-{\rm T})\right]\biggr\}\;,
\end{eqnarray}
 
\begin{eqnarray}\label{dfu3}
\tilde{\Xi}_3(z)&=&
-\frac{15}{2{\rm T}^2}\left[j_2(z)+\frac{9}{2}j_4(z)\right]
\\\nonumber\\\nonumber\mbox{}&&-
\frac{105{\rm h}}{4{\rm T}^3}\biggl\{3\left[j_1(z+{\rm T})-j_1(z-{\rm T})\right]
-7\left[j_3(z+{\rm T})-j_3(z-{\rm T})\right]\biggr\}+\\\nonumber\\\nonumber\mbox{}&&
\frac{105{\rm h}}{4{\rm T}^4}\biggl\{7\left[j_0(z+{\rm T})+ j_0(z-{\rm T})\right]
-50\left[j_2(z+{\rm T})+j_2(z-{\rm T})\right]\biggr\}\;,
\end{eqnarray} 

\begin{eqnarray}\label{dfu4}
\frac{1}{i}\tilde{\Xi}_4(z)&=&\frac{35}{2{\rm T}^3}\left[
j_3(z)+\frac{11}{2}j_5(z)\right]
\\\nonumber\\\nonumber\mbox{}&&-
\frac{315{\rm h}}{4{\rm T}^4}\biggl\{j_0(z+{\rm T})-j_0(z-{\rm T})
-5\left[j_2(z+{\rm T})-j_2(z-{\rm T})\right]\biggr\}\\\nonumber\\\nonumber\mbox{}&&-
\frac{1575{\rm h}}{4{\rm T}^5}\biggl\{9\left[
j_1(z+{\rm T})+ j_1(z-{\rm T})\right]
-28\left[j_3(z+{\rm T})+j_3(z-{\rm T})\right]\biggr\}\;,
\end{eqnarray}

\begin{eqnarray}\label{dfu5}
\tilde{\Xi}_5(z)&=&\frac{315}{8{\rm T}^4}\;j_4(z)
\\\nonumber\\\nonumber\mbox{}&&+
\frac{315{\rm h}}{8{\rm T}^5}\biggl\{3\left[j_1(z+{\rm T})-j_1(z-{\rm T})\right]
-7\left[j_3(z+{\rm T})-j_3(z-{\rm T})\right]\biggr\}\\\nonumber\\\nonumber\mbox{}&&-
\frac{1575{\rm h}}{8{\rm T}^6}\biggl\{\left[j_0(z+{\rm T})+ j_0(z-{\rm T})\right]
-8\left[j_2(z+{\rm T})+j_2(z-{\rm T})\right]\biggr\}\;,
\end{eqnarray}

\begin{eqnarray}\label{dfu6}
\frac{1}{i}\tilde{\Xi}_6(z)&=&-\frac{693}{8{\rm T}^5}\;j_5(z)+
\\\nonumber\\\nonumber\mbox{}&&
\frac{693{\rm h}}{8{\rm T}^6}\biggl\{j_0(z+{\rm T})-j_0(z-{\rm T})
-5\left[j_2(z+{\rm T})-j_2(z-{\rm T})\right]\biggr\}+\\\nonumber\\\nonumber\mbox{}&&
\frac{2079{\rm h}}{8{\rm T}^7}\biggl\{13\left[
j_1(z+{\rm T})+ j_1(z-{\rm T})\right]
-42\left[j_3(z+{\rm T})+j_3(z-{\rm T})\right]\biggr\}\;,
\end{eqnarray}

\begin{eqnarray}\label{dfu7}
\tilde{\Xi}_7(z)&=&j_0(z+{\rm T})+ j_0(z-{\rm T})+\frac{5}{2}
\left[j_2(z+{\rm T})+j_2(z-{\rm T})\right]
\\\nonumber\\\nonumber\mbox{}&&-
\frac{3}{4{\rm T}}\biggl\{3\left[j_1(z+{\rm T})-j_1(z-{\rm T})\right]
-7\left[j_3(z+{\rm T})-j_3(z-{\rm T})\right]\biggr\}\\\nonumber\\\nonumber\mbox{}&&-
\frac{45{\rm h}}{{\rm T}^2}\left[j_2(z)-6j_4(z)\right]\;,
\end{eqnarray}

\begin{eqnarray}\label{dfu8}
\frac{1}{i}\tilde{\Xi}_8(z)&=&
j_0(z+{\rm T})-j_0(z-{\rm T})+
\frac{5}{2}\left[j_2(z+{\rm T})-j_2(z-{\rm T})\right]+
\\\nonumber\\\nonumber\mbox{}&&
\frac{3}{4{\rm T}}\biggl\{3\left[j_1(z+{\rm T})+j_1(z-{\rm T})\right]
-7\left[j_3(z+{\rm T})+j_3(z-{\rm T})\right]\biggr\}+
\\\nonumber\\\nonumber\mbox{}&&
\frac{3{\rm h}}{{\rm T}}
\left[3j_1(z)-7j_3(z)+11j_5(z)\right]\;,
\end{eqnarray}

\begin{eqnarray}\label{dfu9}
\tilde{\Xi}_9(z)&=&\frac{3}{{\rm T}}\biggl\{j_1(z+{\rm T})-j_1(z-{\rm T})+
\frac{7}{2}\left[j_3(z+{\rm T})-j_3(z-{\rm T})\right]\biggr\}
\\\nonumber\\\nonumber\mbox{}&&-
\frac{15}{4{\rm T}^2}\biggl\{\left[j_0(z+{\rm T})+ j_0(z-{\rm T})\right]
-5\left[j_2(z+{\rm T})+j_2(z-{\rm T})\right]\biggr\}
\\\nonumber\\\nonumber\mbox{}&&-
\frac{15{\rm h}}{{\rm T}^2}\left[j_0(z)-5j_2(z)+9j_4(z)\right]\;,
\end{eqnarray}

\begin{eqnarray}\label{dfu10}
\frac{1}{i}\tilde{\Xi}_{10}(z)&=&
-\frac{3}{{\rm T}}\biggl\{\left[j_1(z+{\rm T})+j_1(z-{\rm T})\right]+
\frac{7}{2}\left[j_3(z+{\rm T})+j_3(z-{\rm T})\right]\biggr\}
\\\nonumber\\\nonumber\mbox{}&&-
\frac{15}{4{\rm T}^2}\biggl\{j_0(z+{\rm T})-j_0(z-{\rm T})-
5\left[j_2(z+{\rm T})-j_2(z-{\rm T})\right]+
\\\nonumber\\\nonumber\mbox{}&&
\frac{15{\rm h}}{{\rm T}^3}
\left[-18j_1(z)+77j_3(z)-220j_5(z)\right]\;,
\end{eqnarray}

\begin{eqnarray}\label{dfu11}
\tilde{\Xi}_{11}(z)&=&
-\frac{15}{2{\rm T}^2}\left[j_2(z+{\rm T})+ j_2(z-{\rm T})\right]
\\\nonumber\\\nonumber\mbox{}&&+
\frac{15}{4{\rm T}^3}\biggl\{3\left[j_1(z+{\rm T})-j_1(z-{\rm T})\right]
-7\left[j_3(z+{\rm T})-j_3(z-{\rm T})\right]\biggr\}+\\\nonumber\\\nonumber\mbox{}&&
\frac{15{\rm h}}{{\rm T}^4}
\left[2j_0(z)+5j_2(z)-72j_4(z)\right]\;,
\end{eqnarray}

\begin{eqnarray}\label{dfu12}
\frac{1}{i}\tilde{\Xi}_{12}(z)&=&
-\frac{15}{2{\rm T}^2}\left[j_2(z+{\rm T})- j_2(z-{\rm T})\right]
\\\nonumber\\\nonumber\mbox{}&&-
\frac{15}{4{\rm T}^3}\biggl\{3\left[j_1(z+{\rm T})+j_1(z-{\rm T})\right]
-7\left[j_3(z+{\rm T})+j_3(z-{\rm T})\right]\biggr\}+\\\nonumber\\\nonumber\mbox{}&&
-\frac{15{\rm h}}{{\rm T}^3}
\left[3j_1(z)-7j_3(z)+11j_5(z)\right]\;,
\end{eqnarray}

\begin{eqnarray}\label{dfu13}
\tilde{\Xi}_{13}(z)&=&-
\frac{35}{2{\rm T}^3}\left[j_3(z+{\rm T})- j_3(z-{\rm T})\right]+
\\\nonumber\\\nonumber\mbox{}&&
\frac{35}{4{\rm T}^4}\biggl\{j_0(z+{\rm T})+j_0(z-{\rm T})
-5\left[j_2(z+{\rm T})+j_2(z-{\rm T})\right]\biggr\}+\\\nonumber\\\nonumber\mbox{}&&
\frac{35{\rm h}}{{\rm T}^4}
\left[j_0(z)-5j_2(z)+9j_4(z)\right]\;,
\end{eqnarray}

\begin{eqnarray}\label{dfu14}
\frac{1}{i}\tilde{\Xi}_{14}(z)&=&
\frac{35}{2{\rm T}^3}\left[j_3(z+{\rm T})+ j_3(z-{\rm T})\right]+
\\\nonumber\\\nonumber\mbox{}&&
\frac{35}{4{\rm T}^4}\biggl\{j_0(z+{\rm T})-j_0(z-{\rm T})
-5\left[j_2(z+{\rm T})-j_2(z-{\rm T})\right]\biggr\}+\\\nonumber\\\nonumber\mbox{}&&
\frac{105{\rm h}}{{\rm T}^5}
\left[4j_1(z)-21j_3(z)+66j_5(z)\right]\;.
\end{eqnarray}

\section{Asymptotic behaviour of the dual functions near zero frequency}

In this appendix we give the asymptotic behaviour of the dual functions near zero
frequency. They are as follows:

\begin{eqnarray}\label{dz0}
\tilde{\Xi}_1(f)&=&\tilde{\xi}_1(z)\;,\quad\quad\tilde{\xi}_1(z)=
j_0(z)+\frac{5}{2}j_2(z)+\frac{27}{8}j_4(z)\;,\\\nonumber\\
\frac{1}{i}\tilde{\Xi}_2(f)&=&-\frac{3}{\rm T}\tilde{\xi}_2(z)\;,\quad\quad
\tilde{\xi}_2(z)=j_1(z)+\frac{7}{2}j_3(z)+\frac{55}{8}j_5(z)\;,
\\\nonumber\\
\tilde{\Xi}_3(f)&=&-\frac{15}{2{\rm T}^2}\tilde{\xi}_3(z)\;,\quad\quad
\tilde{\xi}_3(z)=j_2(z)+\frac{9}{2}j_4(z)\;,
\\\nonumber\\
\frac{1}{i}\tilde{\Xi}_4(f)&=&\frac{35}{2{\rm T}^3}\tilde{\xi}_4(z)\;,
\quad\quad\tilde{\xi}_4(z)=j_3(z)+\frac{11}{2}j_5(z)\;,
\\\nonumber\\
\tilde{\Xi}_5(f)&=&\frac{315}{8{\rm T}^4}\tilde{\xi}_5(z)\;,\quad\quad
\tilde{\xi}_5(z)=j_4(z)\;,\\\nonumber\\
\frac{1}{i}\tilde{\Xi}_6(f)&=&-\frac{693}{8{\rm T}^5}\tilde{\xi}_6(z)\;,
\quad\quad\tilde{\xi}_6(z)=j_5(z)\;,\\\nonumber\\
\tilde{\Xi}_7(f)&=&-\frac{45{\rm h}}{{\rm T}^2}\tilde{\xi}_7(z)\;,\quad\quad
\tilde{\xi}_7(z)=j_2(z)-6j_4(z)-\frac{1}{15}z\sin z\;,\\\nonumber\\
\frac{1}{i}\tilde{\Xi}_8(f)&=&\frac{9{\rm h}}{{\rm T}}\tilde{\xi}_8(z)\;,
\quad\quad\tilde{\xi}_8(z)=j_1(z)-\frac{7}{3}j_3(z)+\frac{11}{3}j_5(z)-
\frac{1}{3}\sin z\;,
\\\nonumber\\
\tilde{\Xi}_9(f)&=&-\frac{15{\rm h}}{{\rm T}^2}\tilde{\xi}_9(z)\;,\quad\quad
\tilde{\xi}_9(z)=j_0(z)-5j_2(z)+9j_4(z)-\cos z\;,
\\\nonumber\\
\frac{1}{i}\tilde{\Xi}_{10}(f)&=&-\frac{270{\rm h}}{{\rm T}^3}\tilde{\xi}_{10}(z)\;,
\quad\quad\tilde{\xi}_{10}(z)=j_1(z)-\frac{77}{18}j_3(z)+\frac{110}{9}j_5(z)-
\frac{5}{18}\sin z-\frac{1}{18}z \cos z\;,\\\nonumber\\
\tilde{\Xi}_{11}(f)&=&\frac{30{\rm h}}{{\rm T}^4}\tilde{\xi}_{11}(z)\;,\quad\quad
\tilde{\xi}_{11}(z)=j_0(z)+\frac{5}{2}j_2(z)-36j_4(z)-\frac{1}{2}z \sin z- 
\cos z\;,\\\nonumber\\
\frac{1}{i}\tilde{\Xi}_{12}(f)&=&-\frac{45{\rm h}}{{\rm T}^3}\tilde{\xi}_{12}(z)\;,
\quad\quad\tilde{\xi}_{12}(z)=j_1(z)-\frac{7}{3}j_3(z)+\frac{11}{3}j_5(z)-
\frac{1}{3}\sin z\;,
\\\nonumber\\
\tilde{\Xi}_{13}(f)&=&\frac{35{\rm h}}{{\rm T}^4}\tilde{\xi}_{13}(z)\;,
\quad\quad\tilde{\xi}_{13}(z)=\tilde{\xi}_{9}(z)\;,
\\\nonumber\\\label{dz14}
\frac{1}{i}\tilde{\Xi}_{14}(f)&=&\frac{420{\rm h}}{{\rm
T}^5}\tilde{\xi}_{14}(z)\;,\quad\quad\tilde{\xi}_{14}(z)=
j_1(z)-\frac{63}{12}j_3(z)+\frac{33}{2}j_5(z)-\frac{1}{4}\sin z-
\frac{1}{12}z\cos z\;,\\\nonumber
\\\nonumber
\end{eqnarray}
where ${\rm h}=(-1)^N$.

\section{Asymptotic behaviour of the dual functions near orbital frequency}

The leading terms in the asymptotic expansions of the dual functions $\tilde{\Xi}_a(z)$
near the orbital frequency are as follows:

\begin{eqnarray}\label{dzT}
\tilde{\Xi}_1(f)&=&-\frac{45\rm h}{8{\rm T}}\tilde{\xi}_1(y)\;,\quad\quad
\tilde{\xi}_{1}(y)=j_1(y)-\frac{7}{3}j_3(y)-\frac{1}{3}\sin y\;,\\\nonumber
\\
\frac{1}{i}\tilde{\Xi}_2(f)&=&-\frac{105\rm h}{8{\rm T}^2}\tilde{\xi}_2(y)\;,\quad\quad
\tilde{\xi}_2(y)=j_0(y)-5j_2(y)-\cos y\;,\\\nonumber\\
\tilde{\Xi}_3(f)&=&-\frac{14}{{\rm T}^3}\tilde{\xi}_3(y)\;,\quad\quad
\tilde{\xi}_3(y)=\tilde{\xi}_{1}(y)\\\nonumber
\\
\frac{1}{i}\tilde{\Xi}_4(f)&=&-\frac{6}{{\rm T}^4}\tilde{\xi}_4(y)\;,\quad\quad
\tilde{\xi}_4(y)=\tilde{\xi}_2(y)\;,\\\nonumber \\
\tilde{\Xi}_5(f)&=&\frac{21}{{\rm T}^5}\tilde{\xi}_5(y)\;,\quad\quad
\tilde{\xi}_5(y)=\tilde{\xi}_{1}(y)\;,\\\nonumber\\
\frac{1}{i}\tilde{\Xi}_6(f)&=&\frac{33}{5{\rm T}^6}\tilde{\xi}_6(y)\;,\quad\quad
\tilde{\xi}_6(y)=\tilde{\xi}_2(y)\;,\\\nonumber\\
\tilde{\Xi}_7(f)&=&\tilde{\xi}_7(y)\;,\quad\quad
\tilde{\xi}_7(y)=j_0(y)+\frac{5}{2}j_2(y)\;,\\\nonumber\\
\frac{1}{i}\tilde{\Xi}_8(f)&=&\tilde{\xi}_8(y)\;,\quad\quad
\tilde{\xi}_8(y)=-\tilde{\xi}_7(y)\;,\\\nonumber\\
\tilde{\Xi}_9(f)&=&-\frac{3}{{\rm T}}\tilde{\xi}_9(y)\;,\quad\quad
\tilde{\xi}_9(y)=j_1(y)+\frac{7}{2}j_3(y)\;,\\\nonumber\\
\frac{1}{i}\tilde{\Xi}_{10}(f)&=&\frac{1}{{\rm T}}\tilde{\xi}_{10}(y)\;,\quad\quad
\tilde{\xi}_{10}(y)=\tilde{\xi}_9(y)\;,\\\nonumber\\
\tilde{\Xi}_{11}(f)&=&-\frac{15}{2{\rm T}^2}\tilde{\xi}_{11}(y)\;,\quad\quad
\tilde{\xi}_{11}(y)=j_2(y)\;,\\\nonumber\\
\frac{1}{i}\tilde{\Xi}_{12}(f)&=&\frac{1}{{\rm T}^2}\tilde{\xi}_{12}(y)\;,\quad\quad
\tilde{\xi}_{12}(y)=-\tilde{\xi}_{11}(y)\;,\\\nonumber\\
\tilde{\Xi}_{13}(f)&=&\frac{35}{2{\rm T}^3}\tilde{\xi}_{13}(y)\;,\quad\quad
\tilde{\xi}_{13}(y)=j_3(y)\;,\\\nonumber\\\label{dzTT}
\frac{1}{i}\tilde{\Xi}_{14}(f)&=&\frac{1}{{\rm T}^3}\tilde{\xi}_{14}(y)\;,\quad\quad
\tilde{\xi}_{14}(y)=\tilde{\xi}_{13}(y)\;.\\\nonumber
\end{eqnarray}
where $y=z-{\rm T}$.

\section{Plots of the Fourier transform of autocovariance function}
\clearpage
\begin{figure*}
\centerline{\psfig{figure=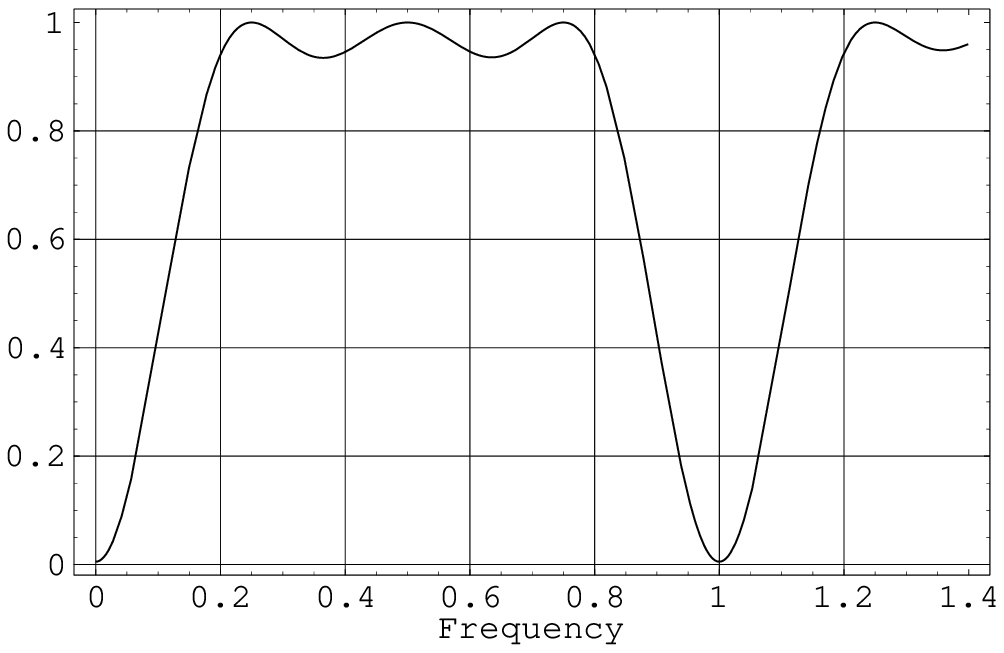,angle=0,height=8cm,width=17cm}}
\caption{Plot of the Fourier transform of  the filter function of 
timing residuals for the amount of
orbital revolutions $N=4$. Frequency is
measured in units of orbital frequency $n_b$. Amplitude of the transform has
been normalized to unity.}
\label{figa1}
\end{figure*}
\begin{figure*}
\centerline{\psfig{figure=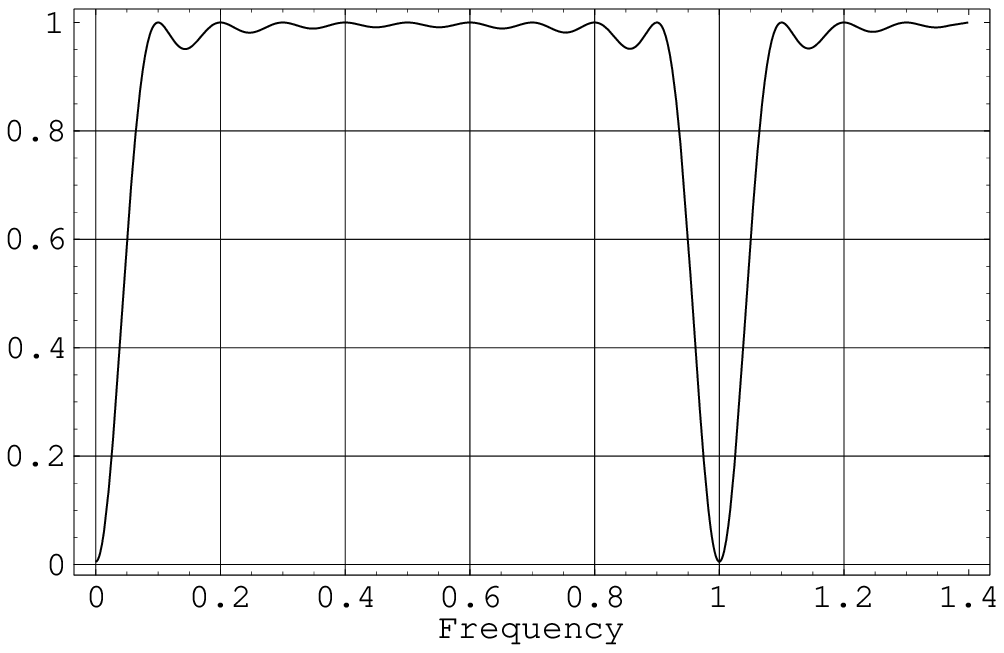,angle=0,height=8cm,width=17cm}}
\caption{Plot of the Fourier transform of  the filter function of 
timing residuals for the amount of
orbital revolutions $N=10$. Frequency is
measured in units of orbital frequency $n_b$. Amplitude of the transform has
been normalized to unity.}
\label{figb1}
\end{figure*}
\begin{figure*}
\centerline{\psfig{figure=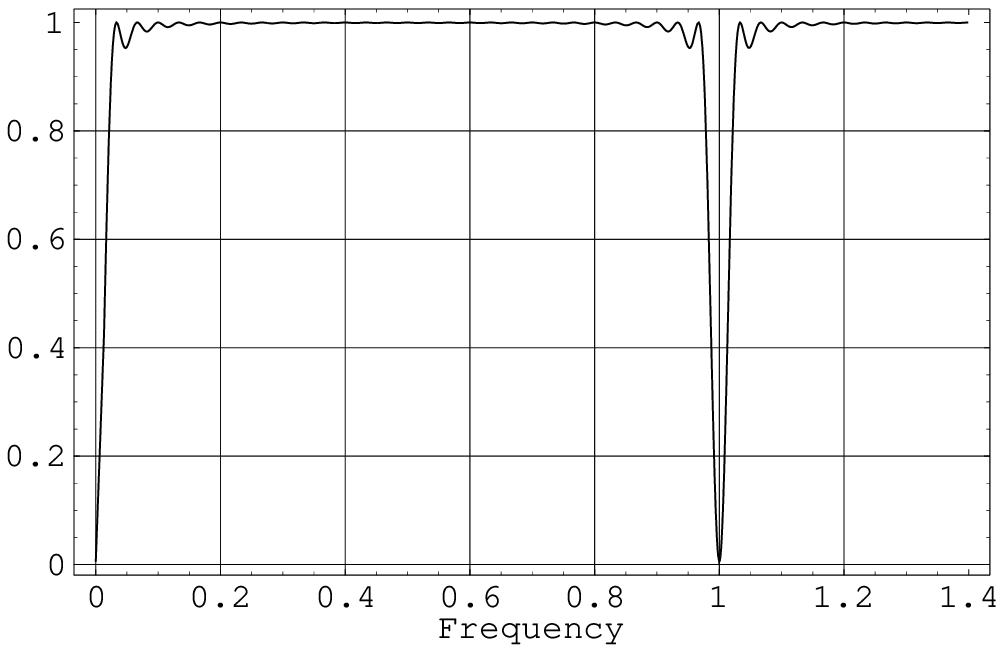,angle=0,height=8cm,width=17cm}}
\caption{Plot of the Fourier transform of the filter function of 
timing residuals for the amount of
orbital revolutions $N=30$. Frequency is
measured in units of orbital frequency $n_b$. Amplitude of the transform has
been normalized to unity.}
\label{figc1}
\end{figure*}

\label{lastpage}

\end{document}